\documentclass[aps,pra,notitlepage,twocolumn,10pt,a4paper]{revtex4-1}

\usepackage{xcolor,graphicx,ulem,soul}
\usepackage{amsmath,amssymb}
\usepackage[colorlinks=true,urlcolor=blue,citecolor=blue,linkcolor=blue]{hyperref}
\usepackage{braket}
\usepackage{subcaption}
\usepackage{dsfont}
\usepackage{comment}

\newcommand{\adag}{\hat{a}^{\dagger}}

\newcommand{\vac}{\ket{\text{vac}}}
\newcommand{\dw}{\text{d}\omega}

\begin{document}

%\preprint{}

\title{A positive operator-valued measure for two-photon detection via sum-frequency generation}


\author{Sofiane Merkouche}
\email[]{Corresponding author: sofianem@uoregon.edu}
\author{Val\'erian Thiel}
\author{Brian J. Smith}
%\homepage[]{}
%\thanks{}
\affiliation{Oregon Center for Optical, Molecular, and Quantum Science, and Department of Physics, University of Oregon, Eugene, OR 97403}


\date{\today}

\begin{abstract}
Spontaneous parametric down conversion (PDC), in the perturbative limit, can be considered as a probabilistic splitting of one input photon into two output photons. Conversely, sum-frequency generation (SFG) implements the reverse process of combining two input photons into one. Here we show that a single-photon projective measurement in the temporal-mode basis of the output photon of a two-photon SFG process effects a generalized measurement on the input two-photon state. We describe the positive operator-valued measure (POVM) associated with such a measurement, and show that its elements are proportional to the two-photon states produced by the time-reversed PDC process. Such a detection acts as a joint measurement on two photons, and is thus an important component of many quantum information processing protocols relying on photonic entanglement. Using the retrodictive approach, we analyze the  properties of the two-photon POVM that are relevant for quantum protocols exploiting two-photon states and measurements.
\end{abstract}

%\keywords{}

\maketitle

%%%%%%%%%%%%%%%%%%%%%%%%%%%%%%%%%%%%%%%%%%%%%%%%%%%%%%%%%%%%%%%%%
\section{Introduction}
Entangled photon pairs are an extremely useful system for studying both the fundamentals \cite{AspectAlainUniversiteParis-Saclay2015} and applications of quantum mechanics, and are the workhorse of experimental quantum optics. This is mainly due to their ease of generation in the laboratory through spontaneous parametric downconversion (PDC), whereby a nonlinear medium such as a crystal is pumped with a bright laser beam and mediates the probabilistic splitting of one pump photon into a pair of photons, subject to energy and momentum conservation. Over the past three decades, much progress has been made in the generation of PDC photon pairs with well-engineered polarization, spectral-temporal, and spatial structure, exhibiting varying degrees of correlation in all of these degrees of freedom. Particular attention has been given recently to encoding quantum information in the spectral-temporal degree of freedom of light. This is because time-frequency modes of light, generally referred to as \textit{temporal modes}, can encode a large amount of information, are particularly well-suited to integrated optics technology, and are robust to communication channel noise \cite{Brecht2015}. In addition, time-frequency entangled photons are useful for applications such as large-alphabet quantum key distribution \cite{Nunn2013}, quantum-enhanced spectroscopy \cite{Raymer2013, Schlawin2017, Dayan2007}, and quantum-enhanced sensing \cite{Zhuang2017}.

Complementary to two-photon state generation is two-photon joint detection, which is an example of the more general concept of a joint quantum measurement on two systems. It is known that joint quantum measurements on separately prepared systems can inherently reveal more information than accessible through separate measurements relying on local operations and classical communication \cite{Bennett1999}. In addition \textit{entangled measurements}, joint measurements whose eigenstates are entangled states, are as crucial a resource as entangled states in quantum protocols such as quantum teleportation \cite{Bennett1993}, remote state preparation \cite{Bennett2001}, entanglement swapping \cite{Halder2007EntanglingTimemeasurement, Sangouard2011FaithfulGeneration}, superdense coding, and quantum illumination \cite{Lloyd1463}. In fact, the equal footing that entangled states and entangled measurements have in quantum protocols such as teleportation has only recently been given due attention \cite{Gisin2019}.

One way to implement a two-photon joint measurement is to use the complement of PDC, sum-frequency generation (SFG). Here two photons interact in a nonlinear medium and are upconverted to a single photon, conserving energy and momentum. Two-photon measurement via SFG has been explored theoretically \cite{Dayan2007} and experimentally \cite{Dayan2005}. In addition, it has been pointed out that the theory of two-photon detection by SFG closely parallels that of two-photon absorption in a molecule, and a unified framework describing both of these processes can be found in reference \cite{Dayan2007}.

In this work we construct and analyze the positive operator valued measure (POVM) associated with joint two-photon measurements relying on SFG followed by mode-selective detection of the upconverted photon in the time-frequency domain. Our development of the two-photon POVM closely parallels that of the POVM for a single photon detected after a filter, as described in reference \cite{VanEnk2017}. We then give some figures of merit for such measurements that are relevant to some of the aforementioned protocols, namely the projectivity, orthogonality, and entanglement of the measurement operators. We illustrate the role of entanglement in measurements with a model of the spectral quantum teleportation scenario. We conclude by highlighting some questions and possible future directions left open by this work.

%%%%%%%%%%%%%%%%%%%%%%%%%%%%%%%%%%%%%%%%%%%%%%%%%%%%%%%%%%%%%%%%%
\section{Framework}

\subsection{The three-wave mixing interaction}
We begin by writing down the transformation describing three-wave mixing, which includes both parametric down-conversion and sum-frequency generation, in the interaction picture. We assume a given polarization configuration and assume that all the interacting fields occupy a single transverse spatial mode, so that only the time-frequency degrees of freedom of the field are relevant. Under these conditions the transformation may be expressed as

\begin{equation}
    \label{hamtwm}
    \begin{gathered}
    \hat{H} = \hat{H}_{PDC}  + \hat{H}_{SFG},\\
    \hat{H}_{PDC}=\chi\int \dw_s \dw_i \Phi(\omega_s,\omega_i)\ \hat{a}_p(\omega_s+\omega_i)\adag_s(\omega_s)\adag_i(\omega_i),\\
    \hat{H}_{SFG} = (\hat{H}_{PDC})^\dagger,
    \end{gathered}
\end{equation}
where $\hat{a}^{(\dagger)}_j(\omega_j)$ is the annihilation (creation) operator for a single photon at monochromatic mode $j$ with frequency $\omega_j$, and $j=p,s,i$ label the pump, signal, and idler frequencies; $\chi \ll 1$ is a parameter characterizing the efficiency of the process, describing the second-order nonlinearity and containing all the parameters that are constant or slowly-varying over the integration; and $\Phi(\omega_s,\omega_i)$ is the \textit{phase-matching function}, which has the form

\begin{equation}
%\Phi(\omega_s, \omega_i)= e^{i \frac{\Delta\mathbf{k}\cdot\mathbf{L}}{2}} \mathrm{sinc}\left(\frac{\Delta\mathbf{k}\cdot\mathbf{L}}{2}\right),
\Phi(\omega_s, \omega_i) \propto  \mathrm{sinc}\left(\frac{\Delta\mathbf{k}\cdot\mathbf{L}}{2}\right),
\end{equation}
where $\mathbf{L}$ is the vector quantifying the length of the interaction medium, and $\Delta\mathbf{k}=\mathbf{k}_p(\omega_s+\omega_i)-\mathbf{k}_s(\omega_s)-\mathbf{k}_i(\omega_i)$ is the wavevector mismatch for the three fields. $\Phi$ takes on its maximum value when $\Delta\mathbf{k}=\mathbf{0}$, and thus corresponds to momentum conservation in the process. Finally, we have separated the transformation explicitly into $\hat{H}_{PDC}$, the term responsible for PDC, and its Hermitian conjugate, $\hat{H}_{SFG}$, responsible for SFG.

The interacting fields evolve unitarily under this transformation, and for our analysis, we will consider only the weak-interaction limit, so that, for an input state $\ket{\Psi_\text{in}}$, the output state is given by
\begin{equation}
\label{hint}
    \ket{\Psi_\text{out}} = \exp{[-i \hat{H}]}\ket{\Psi_\text{in}} \approx \left(1 - i\hat{H}\right)\ket{\Psi_\text{in}}.
\end{equation}
Note that, in a slight abuse of notation, we are using $\hat{H}$ to reflect the fact that this transformation is derived from the interaction Hamiltonian for three-wave mixing, although the latter is a time-dependent quantity with a different dimensionality (see Appendix \ref{app:ham}).

\subsection{PDC photon pairs and the joint spectral amplitude}

\begin{figure}
    \includegraphics[width=6cm]{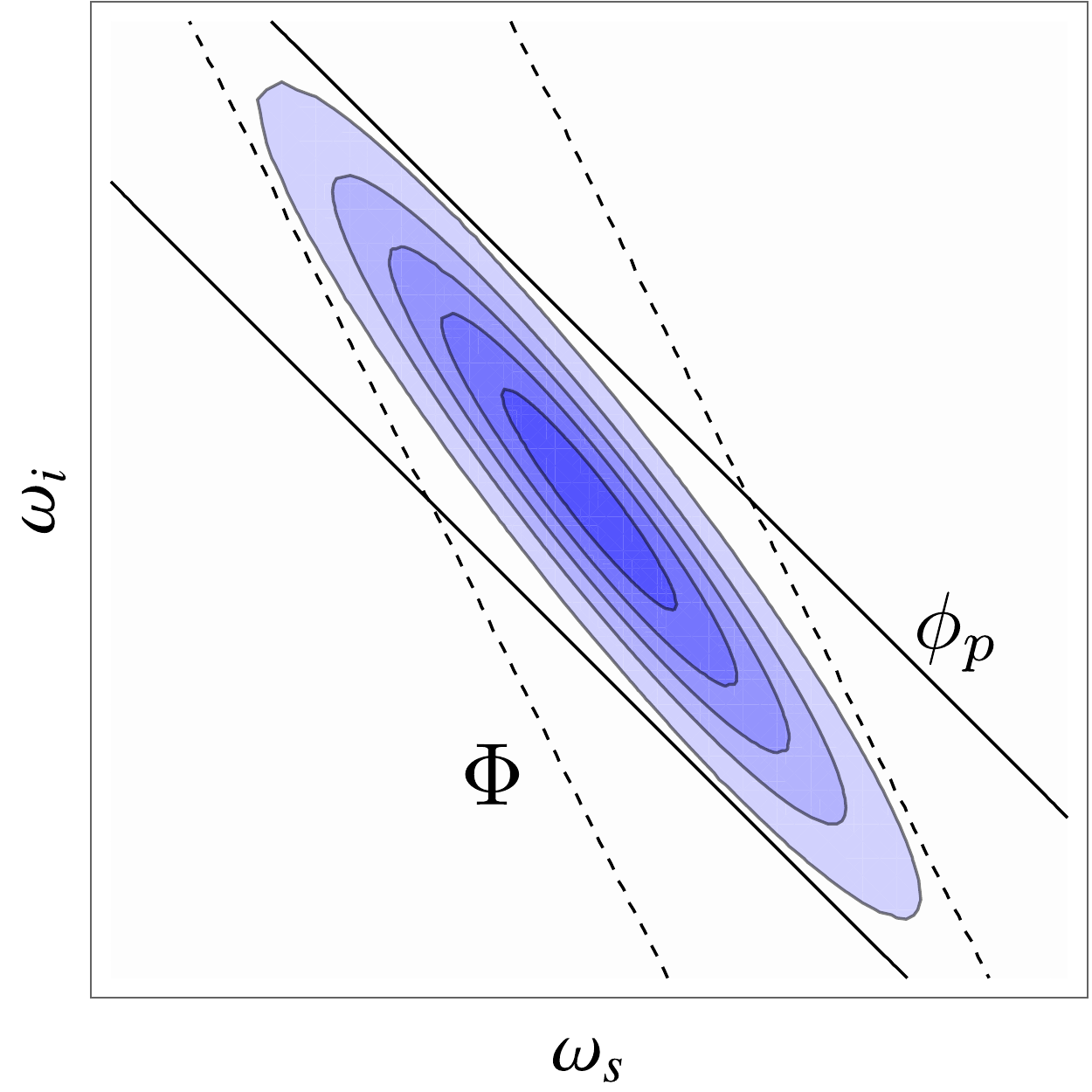}
    \caption{Two-dimensional plot of the magnitude of a typical JSA. The solid lines contour a Gaussian pump mode $\phi_p(\omega_s+\omega_i)$, and the dashed lines contour the phasematching function $\Phi(\omega_s,\omega_i)$. This shows how spectral correlations arise in the JSA. Frequencies are in arbitrary units.}
    \label{jsafig}
\end{figure}

It is instructive to briefly review the spectral-temporal structure of photon pairs generated by PDC, governed by the $\hat{H}_{PDC}$ term. In most applications PDC is pumped by a strong coherent state occupying a spectral mode function $\phi_p(\omega)$, which can be treated as a classical field amplitude $E_p(\omega)=E_0 \phi_p(\omega)$, where $E_0$ quantifies the field strength, and $\phi_p(\omega)$ is normalized as $\int \dw\ |\phi_p(\omega)|^2=1$. However, since we are working in the perturbative limit, it is equivalent to consider a single-photon pump in the state

\begin{equation}
    \ket{\Psi_\text{in}} = \ket{\phi_p}=\int \dw \phi_p(\omega) \hat{a}_p^{\dagger}(\omega)\ket{\text{vac}}.
\end{equation}
After this state undergoes unitary evolution according to equation \eqref{hint}, we obtain the output state

\begin{equation}
\label{psiout}
    \ket{\Psi_\text{out}}=\ket{\phi_p}-i\sqrt{w}\ket{\Psi_\text{PDC}},
\end{equation}
where
\begin{equation}
    \ket{\Psi_\text{PDC}}=\frac{\chi}{\sqrt{w}}\int \dw_s \dw_i \phi_p(\omega_s+\omega_i)\Phi(\omega_s,\omega_i)\hat{a}^{\dagger}_s(\omega_s)\hat{a}^{\dagger}_i(\omega_i)\ket{\text{vac}}
\end{equation}
is a normalized two-photon state, and where
\begin{equation}
    w=\int \dw_s \dw_i |\chi\ \phi_p(\omega_s+\omega_i)\Phi(\omega_s,\omega_i)|^2
\end{equation}
is a normalization factor. 

It is convenient here to define the \textit{joint spectral amplitude} (JSA)

\begin{equation}
f(\omega_s,\omega_i) = \frac{\chi}{\sqrt{w}}\phi_p(\omega_s+\omega_i) \Phi(\omega_s,\omega_i),
\end{equation}
so that

\begin{equation}
    \ket{\Psi_\text{PDC}}=\int \dw_s \dw_i f(\omega_s,\omega_i) \adag_s(\omega_s) \adag_i(\omega_i)\vac
\end{equation}
The JSA can be viewed as a two-photon wavefunction, and its modulus squared, $|f(\omega_s,\omega_i)|^2$, is the probability density function for the photon pair in frequency space, normalized as $\int \dw_s\dw_i |f(\omega_s,\omega_i)|^2=1$. Considerable progress has been made in engineering the temporal-mode structure of PDC photon pairs, which is completely characterized by the JSA, and this is done by shaping of the pump spectral amplitude $\phi_p(\omega_s+\omega_i)$ and engineering of the phasematching $\Phi(\omega_s,\omega_i)$ in the nonlinear medium. We plot schematically in Fig. \ref{jsafig} a typical JSA configuration showing its dependence on the pump amplitude and the phasematching function. A thorough review of the state-of-the-art in two-photon state engineering in the time-frequency domain can be found in reference \cite{Ansari2018}.

\subsection{Two-photon SFG and the two-photon POVM}

\begin{figure}
\includegraphics[width=6cm]{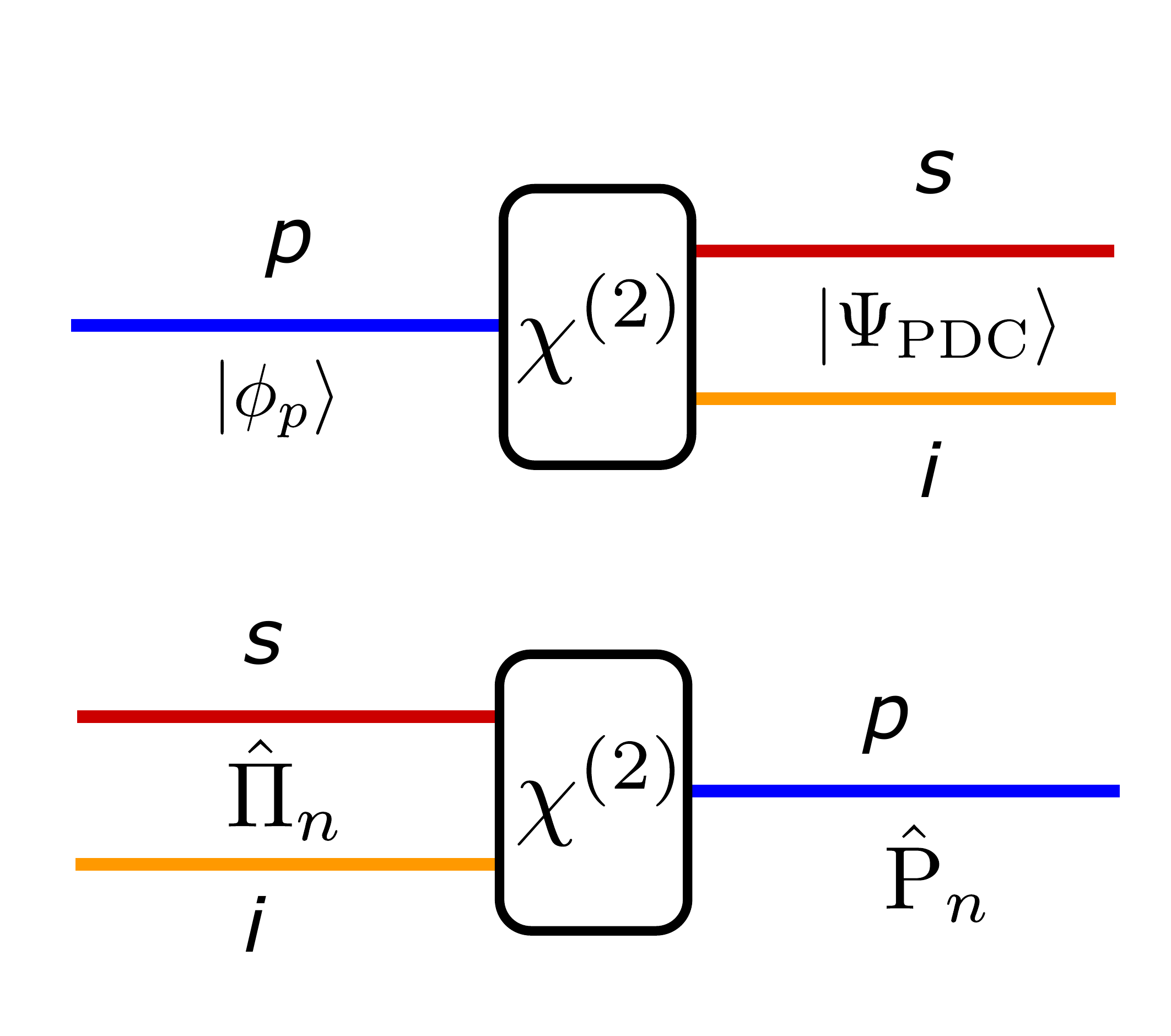}
\caption{PDC uses a $\chi^{(2)}$ interaction medium to convert a single-photon state $\ket{1_\phi}$ in the mode $p$ to a pair of photons in modes $s$ and $i$, described by the state $\ket{\psi_\text{PDC}}$ given in the text. In the time-reverse picture, a projective measurement $\hat{\text{P}}_n$ of a single photon produced by SFG implements measurement with POVM element $\hat{\Pi}_n$ on the two input photons.}
\label{figsch}
\end{figure}

We now turn our attention to the SFG term in equation \eqref{hamtwm}, explicitly given by

\begin{equation}
    \hat{H}_{SFG} = \chi^*\int \dw_s \dw_i \Phi^*(\omega_s,\omega_i)\hat{a}^{\dagger}_p(\omega_s+\omega_i)\hat{a}_s(\omega_s)\hat{a}_i(\omega_i)
\end{equation}
and consider the upconversion of an arbitrary pure two photon state given by 

\begin{equation}
\ket{\Psi_\text{in}}=\ket{\psi_g}=\int \dw_s \dw_i g(\omega_s,\omega_i) \adag_s(\omega_s)\adag_i(\omega_i)\vac,
\end{equation}
where $g(\omega_s,\omega_i)$ is a two-photon JSA. The output state will then be

\begin{equation}
    \ket{\Psi_\text{out}}= \ket{\psi_g} - i \chi^*\ket{\sigma},
\end{equation}
where
\begin{equation}
    \ket{\sigma} = \int d\nu \sigma(\nu) \adag_p(\nu)\vac,
\end{equation}
with the (unnormalized) spectral amplitude function
\begin{equation}
\label{sig}
    \sigma(\nu) = -\frac{1}{2}\int d\nu'\ \tilde{\Phi}^*\left(\nu,\nu'\right) \tilde{g}\left(\nu,\nu'\right).
\end{equation}
We obtain this last equation by changing variables to the sum and difference frequencies $\nu=\omega_s+\omega_i$ and $\nu'=\omega_s-\omega_i$, and defining $\tilde{\Phi}^*(\nu,\nu') = \Phi^*\left(\frac{\nu+\nu'}{2},\frac{\nu-\nu'}{2}\right)$ (and likewise for $\tilde{g}(\nu,\nu')$).

We are now equipped to develop the two-photon POVM corresponding to a detection of the upconverted single-photon state $\ket{\sigma}$, which closely mirrors the one-photon, pre-filter POVM described in reference \cite{VanEnk2017}. Consider performing an ideal, projective measurement of the upconverted photon onto an orthonormal set of temporal mode single photon states $\{(\hat{\text{P}}_n=\ket{\phi_n}\bra{\phi_n})_{n=1}^\infty\}$ with 
\begin{equation}
   \ket{\phi_n}=\int \dw \phi_n(\omega)\adag_p(\omega)\vac, 
\end{equation}

satisfying

\begin{equation}
    \braket{\phi_n|\phi_m}=\int \dw\ \phi^*_n(\omega)\phi_m(\omega)=\delta_{nm}.
\end{equation}
Such a measurement can in principle be realized using a quantum pulse gate, recently described and demonstrated in references \cite{Ansari2018_prl, Reddy2018}, whereby a strong pump field in a particular temporal mode selects out that same mode from an input signal field and upconverts it through SFG to a register mode which can be easily detected with a spectrometer. The probability for a successful detection for this measurement will be given by

\begin{equation}
\begin{gathered}
    p_n= |\chi^*\braket{\phi_n|\sigma}|^2\\
   =\left|-\frac{\chi^*}{2}\int\ \text{d}\nu \text{d}\nu'\ \phi_n^*(\nu) \tilde{\Phi}^*\left(\nu,
   \nu'\right) \tilde{g}\left(\nu,\nu'\right)\right|^2\\
    = \left|\chi^*\int \dw_s \dw_i \phi_n^*(\omega_s+\omega_i)\Phi^*(\omega_s,\omega_i)g(\omega_s,\omega_i)\right|^2
\end{gathered}    
\end{equation}
However, this same probability can be obtained by applying the Born rule to the input state $\hat{\rho}_\text{in}=\ket{\Psi_\text{in}}\bra{\Psi_\text{in}}$ in the two-photon space:

\begin{equation}
p_n=\text{Tr}(\hat{\rho}_\text{in}\hat{\Pi}_n),
\end{equation}
if we define a POVM element
\begin{equation}
    \hat{\Pi}_n = w_n \ket{\Psi_n}\bra{\Psi_n},
\end{equation}
where

\begin{equation}
    \ket{\Psi_n}=\frac{\chi}{\sqrt{w_n}}\int \dw \dw' \phi_n(\omega+\omega')\Phi(\omega,\omega')\adag_s(\omega)\adag_i(\omega')\vac,
\end{equation}
and

\begin{equation}
    w_n =  \int \dw \dw' |\chi\ \phi_n(\omega+\omega')\Phi(\omega,\omega')|^2.
\end{equation}

We immediately recognize $\ket{\Psi_n}$ as the normalized two-photon state that would result from PDC with a pump photon in the state $\ket{\phi_n}$. That is, a projective measurement of an upconverted photon with projector $\hat{\text{P}}_n=\ket{\phi_n}\bra{\phi_n}$ implements a generalized measurement of the two input photons with POVM element $\hat{\Pi}_n$. This is schematically shown in Fig. \ref{figsch}. Furthermore, the properties of $\hat{\Pi}_n$ follow immediately from the properties of the PDC state $\ket{\Psi_n}$, as we will see in the following section. It is convenient to associate with the POVM element $\hat{\Pi}_n$ a \textit{measurement} JSA 

\begin{equation}
    f_n(\omega+\omega')=\frac{\chi}{\sqrt{w_n}}\phi_n(\omega+\omega')\Phi(\omega,\omega').
\end{equation}

\begin{comment}
To complete the POVM, we note that the probability for no detection, which is due to no upconversion since we assume an ideal detector, is 

\begin{equation}
    p_\text{null} = 1 - \sum_{n=1}^\infty p_n = 1 - |\chi|^2,
\end{equation}
and we can define a corresponding POVM element for no detection as

\begin{equation}
    \hat{\Pi}_\text{null}=\mathds{1} - \sum_{n=1}^\infty \hat{\Pi}_n,
\end{equation}
where $\mathds{1}$ denotes the identity operator in the relevant two-photon subspace. 

\end{comment}

To complete the POVM, we note that we are considering an ideal detector in the SFG mode, such that any upconverted photon is detected with certainty. We are thus justified in defining an element corresponding to no detection as

\begin{equation}
    \hat{\Pi}_\text{null}=\mathds{1} - \sum_{n=1}^\infty \hat{\Pi}_n,
\end{equation}
where $\mathds{1}$ denotes the identity operator in the relevant two-photon subspace. Using the fact that the $\phi_n$ mode functions form a complete orthonormal set, we can evaluate

\begin{equation}
     \sum_{n=1}^\infty \hat{\Pi}_n = |\chi|^2 \int \dw \dw' |\Phi(\omega,\omega')|^2 \ket{\omega,\omega'}\bra{\omega,\omega'},
\end{equation}
where $\ket{\omega,\omega'}=\adag_s(\omega)\adag_i(\omega_i)\vac$. Noting that the identity in the two-photon subspace can be resolved as

\begin{equation}
     \mathds{1} = \int \dw \dw' \ket{\omega,\omega'}\bra{\omega,\omega'},
\end{equation}
we can express $\hat{\Pi}_\text{null}$ explicitly as

\begin{equation}
	\hat{\Pi}_\text{null}= \int \dw \dw' \left(1-|\chi|^2 |\Phi(\omega,\omega')|^2\right) \ket{\omega,\omega'}\bra{\omega,\omega'}.
\end{equation}
Finally we may write down the complete two-photon POVM as

\begin{equation}
    \left\{(\hat{\Pi}_n)_{n=1}^\infty,\hat{\Pi}_\text{null}\right\},
\end{equation}
satisfying

\begin{equation}
    \sum_{n=1}^\infty \hat{\Pi}_n + \hat{\Pi}_\text{null} = \mathds{1}.
\end{equation}

%%%%%%%%%%%%%%%%%%%%%%%%%%%%%%%%%%%%%%%%%%%%%%%%%%%%
\section{Properties of the measurement operator}

\subsection{Projectivity}
We will now take advantage of the well-studied properties of the two-photon PDC state $\ket{\Psi_n}$ to analyze some of the useful properties of the POVM element $\hat{\Pi}_n$. We begin by defining the retrodicted two-photon state \cite{Amri2011}, corresponding to an outcome $n$, as

\begin{equation}
\label{puretr}
    \hat{\rho}_n = \frac{\hat{\Pi}_n}{\text{Tr}(\hat{\Pi}_n)} = \ket{\Psi_n}\bra{\Psi_n}.
\end{equation}
We consider the measurement \textit{projective}, if $\hat{\rho}_n$ is a pure state, satisfying $\text{Tr}(\hat{\rho}_n^2)=1$, which is indeed the case for equation \eqref{puretr}. 

In general, however, single-photon detectors are not perfectly resolving. In the case of the quantum pulse gate, a detector click may not correspond to single pulse mode, but rather an incoherent mixture of a few modes. In the case of a non-ideal spectrally resolving detection, one either uses a filter of finite bandwidth, or a spectrometer with finite resolution. In all of these cases, it is more accurate to describe a non-ideally resolving, that is, non-projective, single-photon measurement by 

\begin{equation}
    \hat{\text{P}}_q= \sum_{n} q_n \hat{\text{P}}_n
\end{equation} where $0\leq q_n \leq 1$ are weighting coefficients. This leads to a two-photon POVM element

\begin{equation}
    \hat{\Pi}_q = \sum_n q_n \hat{\Pi}_n,
\end{equation}
and a retrodicted state

\begin{equation}
    \hat{\rho}_q=\frac{\hat{\Pi}_q}{\text{Tr}(\hat{\Pi}_q)},
\end{equation}
which has $\text{Tr}(\hat{\rho}_q^2)\leq 1$ and is not in general a pure state. Evidently, the two-photon POVM elements are projective if and only if the single-photon measurement operators are projective.

Projective two-photon measurements are of particular importance in quantum teleportation and remote-state preparation, and entanglement swapping, because in these schemes the measurement acts as a herald to a single photon state or a two-photon entangled state, respectively. Ideally the heralded states should be pure to be useful for quantum information processing. And the purity of the heralded state is limited by both the purity of the input states and the purity (projectivity) of the heralding measurement \cite{Amri2011}.

\subsection{Orthogonality}
Orthogonal measurements are measurements which project onto orthogonal states, and thus satisfy

\begin{equation}
\hat{\Pi}_n\hat{\Pi}_m \propto \delta_{nm}\hat{\Pi}_n.
\end{equation}
We note here that orthogonal measurements of the SFG photon do not correspond to orthogonal two-photon POVM elements in general. This is analogous to the fact that PDC pumped with orthogonal pulse modes does not produce orthogonal PDC states in general. The non-orthogonality of the two-photon states can be seen by taking

\begin{equation}
\begin{gathered}
\label{overlap}
    \braket{\Psi_n|\Psi_m}=\\
    \frac{|\chi|^2}{\sqrt{w_nw_m}}\int \dw \dw' \phi^*_n(\omega+\omega')\phi_m(\omega+\omega')|\Phi(\omega,\omega')|^2 \neq \delta_{nm}.
\end{gathered}
\end{equation}
This is due to the filtering induced by the phasematching function. This is indeed analogous to what happens when two orthogonal modes are subjected to linear filtering (see reference \cite{VanEnk2017} on this point): in general the transmitted modes considered alone are not orthogonal, even though filtering is a unitary process. The orthogonality is preserved only when considering all of the modes involved in the transformation, whereas here we are only considering the signal and idler modes and not the pump.

\begin{figure}[t]
  \includegraphics[width=9cm]{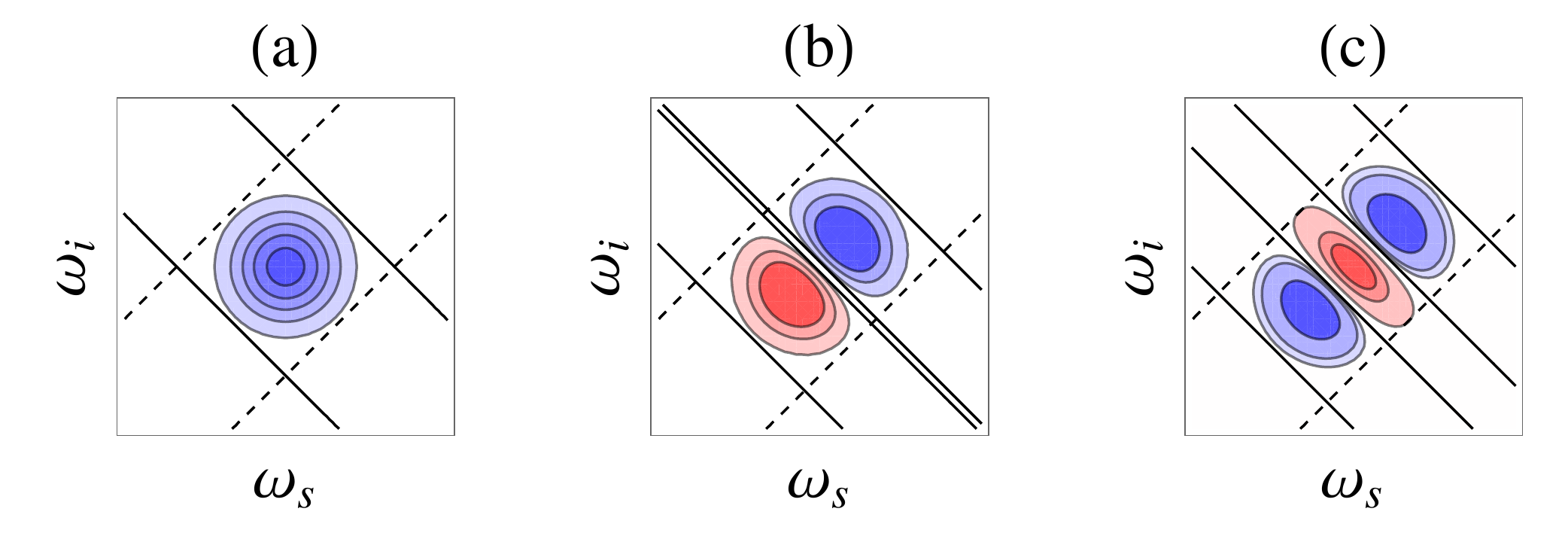}
  \caption{JSA's for the configuration described in the text where the phasematching function is engineered through group-velocity matching makes an angle $\theta = 45^\text{o}$ with respect to the $\omega_s$-axis. Then it becomes independent of the sum frequency $\nu = \omega_s+\omega_i$, and thus orthogonal measurements of the SFG photon correspond to orthogonal two-photon POVM elements. Blue (red) indicates positive (negative) amplitudes. In the case of PDC, the amount of correlations in the JSA can be controlled by shaping of the pump pulse, as described in reference \cite{Ansari2018}. Here we plot the JSA's obtained by shaping the pump into the (a) zeroth-, (b) first-, and (c) second-order Hermite-Gauss modes, resulting into mutually-orthogonal two-photon states. Frequencies are in arbitrary units.}
  \label{fig:pm}
\end{figure}

An obvious question that arises then is, in what cases do the POVM elements, in fact, correspond to orthogonal measurements? The answer to this question becomes obvious when we rewrite equation \eqref{overlap} in terms of the sum and difference frequencies $\nu$ and $\nu'$,

\begin{equation}
\begin{gathered}
    \braket{\Psi_n|\Psi_m}=\\
    \frac{|\chi|^2}{4\sqrt{w_nw_m}}\int \text{d}\nu \text{d}\nu' \phi^*_n(\nu)\phi_m(\nu)\left|\tilde{\Phi}\left(\nu,\nu'\right)\right|^2.
\end{gathered}
\end{equation}
Clearly, only when the phasematching function does not depend on the sum-frequency $\nu$, that is, $\Phi=\Phi(\nu')$, then do we obtain

\begin{equation}
\label{ortho}
    \braket{\Psi_n|\Psi_m}=\delta_{nm},
\end{equation}
and the $\hat{\Pi}_n$ then satisfy
\begin{equation}
    \hat{\Pi}_n\hat{\Pi}_m=\delta_{nm}w_n\hat{\Pi}_n.
\end{equation}

Orthogonality of the two-photon POVM elements is of interest, for example, in the quantum illumination scheme as originally described by Lloyd \cite{Lloyd1463}. Here an entangled two-photon state $\ket{\Psi_n}$ is prepared and one of the photons sent to reflect off a possibly present target, while the other photon is kept in the lab. The two photons are then to be jointly measured, whereupon a successful projection onto the initial state $\ket{\Psi_n}$ indicates the presence of the target. If one is to implement this scheme using SFG as the two-photon measurement, non-orthogonal measurements would suffer from the possibility that the desired state $\ket{\Psi_n}$ could give a positive outcome corresponding to the ``wrong" measurement associated with a non-orthogonal state $\ket{\Psi_m}$.

In general, the orthogonality condition \eqref{ortho} can be approximately satisfied as long as the phase-matching function varies slowly enough in the $\nu$ direction, in comparison to the support of the detection mode function. This happens, for example, in a sufficiently short interaction medium. However, there are two limiting cases that are of note. The first is the spectrally resolved detection limit, which corresponds to simply measuring the output with an ideal spectrometer. In this limit, the detection mode can be approximated by a delta function,

\begin{equation}
    \phi_n(\omega) \rightarrow \delta(\omega-\omega_n),
\end{equation}
and
\begin{equation}
    f_n(\omega,\omega') \propto \delta(\omega+\omega'-\omega_n),
\end{equation}
where $\omega_n$ is the measured frequency at the spectrometer. This is the analogue of pumping a PDC source with monochromatic, or continuous-wave (cw), light. In both of these cases, orthogonal pump (or measurement modes) with frequencies $\omega_n$ and $\omega_m$ correspond to orthogonal two-photon states (or measurements) with sum frequencies $\omega_n$ and $\omega_m$. 

The second case of interest is achieved by extended phase-matching techniques, as described in reference \cite{Ansari2018}. For certain nonlinear materials and field configurations, it is possible, using group-velocity matching, to make the phase-matching function approximately constant in the $\nu$ direction over some range of interest. More precisely, the phase-matching function can be engineered to make an angle $\theta = 45^\text{o}$ in the $\omega_s$-$\omega_i$ plane, perpendicular to the angle that the pump function makes. This configuration has been used by Ansari \textit{et al} to generate PDC states with a controllable temporal-mode structure and degree of entanglement through pump pulse-shaping \cite{Ansari2018_prl}. This concept is illustrated schematically in Fig. \ref{fig:pm}. More recently, similarly exotic two-photon states have been obtained through phasematching shaped by the periodic poling of the nonlinear crystal, rather than pulse-shaping of the pump \cite{Graffitti2020}.

An interesting result that follows from the limit where $\Phi$ is independent of $\nu$ is the possibility of downconverting an arbitrary pulse shape in a nonlinear medium into an entangled photon pair, and recovering the pump pulse shape by upconverting the photon pair in an identical medium. This can be seen by taking $\tilde{g}(\nu,\nu')=\phi(\nu)\tilde{\Phi}(\nu')$ in equation \eqref{sig}, and obtaining

\begin{equation}
    \sigma(\nu)=\phi(\nu)\int d\nu' |\tilde{\Phi}(\nu')|^2,
\end{equation}
which is evidently proportional to the input $\phi(\nu)$. The spatial analogue of this result, whereby a pump beam shaped in a specific transverse spatial mode is downconverted, and the photon resulting from the upconversion of the PDC pair is shown to recover the transverse spatial mode, has recently been experimentally demonstrated by Jimenez \textit{et al} \cite{Jimenez2019}. 

%%%%%%%%%%%%%%%%%
\subsection{Entanglement} \label{sec:entanglement}

We now turn to perhaps a more interesting question regarding the two-photon measurement operator: when is the POVM element $\hat{\Pi}_n$ a projector onto an entangled two-photon state, and thus can be said to enact an \textit{entangled measurement} on the input photons? \cite{Vertesi2011,Renou2018} We can answer this question readily: $\hat{\Pi}_n$ is an entangled measurement, if the retrodicted state $\rho_n$ is an entangled state. Entangled measurements play a central role in quantum teleportation, superdense coding, and quantum illumination, among many other protocols, and recently the role of entanglement in joint measurements has been recognized to be equally important to the role of entanglement of states as a shared resource \cite{Gisin2019}.

To illustrate the role of entangled measurements in a quantum protocol, we will investigate briefly the spectral quantum teleportation scenario, described by Molotkov \cite{Molotkov1998} and by Humble \cite{Humble2010} (and whose spatial analogue was described by Walborn \textit{et al} \cite{Walborn2007}). In this protocol, Alice and Bob share a two-photon entangled state described by a JSA $f_s(\omega_a,\omega_b)$, and Alice is to teleport a single photon state with spectral amplitude $\psi_c(\omega_c)$ by performing an SFG measurement on this photon and her half of the entangled state, and communicating the measurement result to Bob.

\begin{figure}[t]
  \includegraphics[width=8cm]{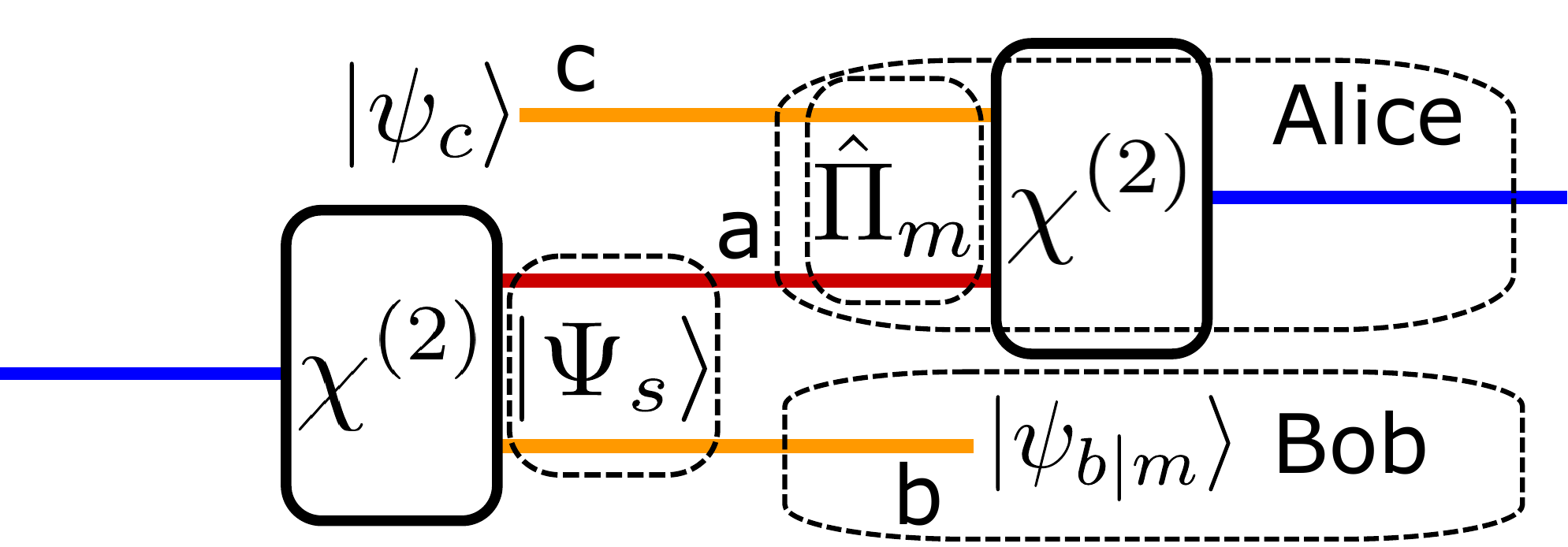}
  \caption{Spectral teleportation scenario considered in the text. Alice and Bob share entangled photons $a$ and $b$ in the state $\ket{\Psi_s}$. Alice performs a two-photon SFG measurement $\hat{\Pi}_m$ on her photon $a$ and photon $c$, in the state $\ket{\psi_c}$, and communicates the result of her measurement to Bob, whereupon Bob reconstructs the state $\ket{\psi_{b|m}}$.}
   \label{fig:telep}
\end{figure}

Reference \cite{Molotkov1998} considers only the case of a maximally-correlated pair of entangled photons shared between Alice and Bob, while reference \cite{Humble2010} generalizes this result to the case of a Gaussian JSA, which is a good approximation to what can be produced using pulsed lasers as a pump. In both references however, Alice's joint measurement is a spectrally-resolved measurement of the SFG photon. Here we use our formalism to generalize further to a pulse-mode resolved measurement of the SFG photon, as can be realized with a quantum pulse gate, by considering a generalized measurement JSA $f_m(\omega_a, \omega_c)$. It was first pointed out in the original proposal of quantum teleportation \cite{Bennett1993} that in addition to the maximally-entangled state (generalized Bell-state) shared by Alice and Bob, quantum teleportation with unit fidelity is achieved when Alice's joint measurement projects onto a maximally-entangled state. Here we show behavior that is consistent with this result by quantifying the teleportation fidelity as a function of the entanglement of \textit{both} the shared state \textit{and} the joint measurement. It is worth clarifying that our current goal is not to \textit{demonstrate} that the POVM element is entangled, but rather, it is to show that our POVM formalism is sufficient to describe quantum teleportation in the time-frequency domain, provided we stipulate entanglement as a property of the measurement. This is in keeping with the more familiar case of the Bell-state measurement's role in qubit teleportation.

The teleportation scenario we consider is shown schematically in Fig. \ref{fig:telep}. Alice and Bob share entangled photons a and b, respectively, described by a Gaussian JSA similar to the one in reference \cite{Humble2010}:

\begin{equation}
	\begin{gathered}
    	\ket{\Psi_s}=\int \dw_a\ \dw_b\ f_s(\omega_a,\omega_b) \adag_a(\omega_a)\adag_b(\omega_b) \vac\\
    	f_s(\omega_a,\omega_b) = N_s\text{Exp}\left[-\frac{1}{\gamma_s^2(1-\alpha^2)}\left(\frac{\omega_a^2}{2}+\frac{\omega_b^2}{2}+\alpha \omega_a\omega_b\right)\right]
	\label{eq:fstate}
	\end{gathered}
\end{equation}
where $\alpha \in [-1,1]$ is the correlation between the the photon frequencies, with $\alpha=1$ corresponding to maximal frequency anticorrelation, such as would be obtained from a cw pump; $\gamma_s$ is the characteristic bandwidth of the PDC photons, and $N_s$ is the normalization constant. Alice provides a single photon c to be teleported, described by the state

\begin{equation}
    \ket{\psi_c} = \int \dw_c \psi_c(\omega_c) \adag_c(\omega_c)\vac
\end{equation}
where $\psi_c(\omega_c)$ is an arbitrary spectral amplitude function. Alice initiates the teleportation by performing an SFG measurement on photons a and c, represented by an operator $\hat{\Pi}_m = w_m\ket{\Psi_m}\bra{\Psi_m}$, with

\begin{equation}
\begin{gathered}
    \ket{\Psi_m}=\int \dw_a\ \dw_c\ f_m(\omega_a,\omega_c) \adag_a(\omega_a)\adag_c(\omega_c) \vac\\
    f_m(\omega_a,\omega_c) = N_m\text{Exp}\left[-\frac{1}{\gamma_m^2(1-\beta^2)}\left(\frac{\omega_a^2}{2}+\frac{\omega_c^2}{2}+\beta \omega_a\omega_c\right)\right]
\end{gathered}
\end{equation}
with parameters defined similarly to $\ket{\Psi_s}$. 

%To simplify our analysis so that the only variable parameters are the correlations , we take $f_s$ and $f_m$ to be centered at identical frequencies (here 0 for simplicity) and to have the same characteristic widths in each direction. Experimentally this corresponds to using identical crystals for the down conversion and up conversion, and conditioning the teleportation on up conversion events centered at the PDC pump frequency.

\begin{figure*}[t]
  \includegraphics[width=\textwidth]{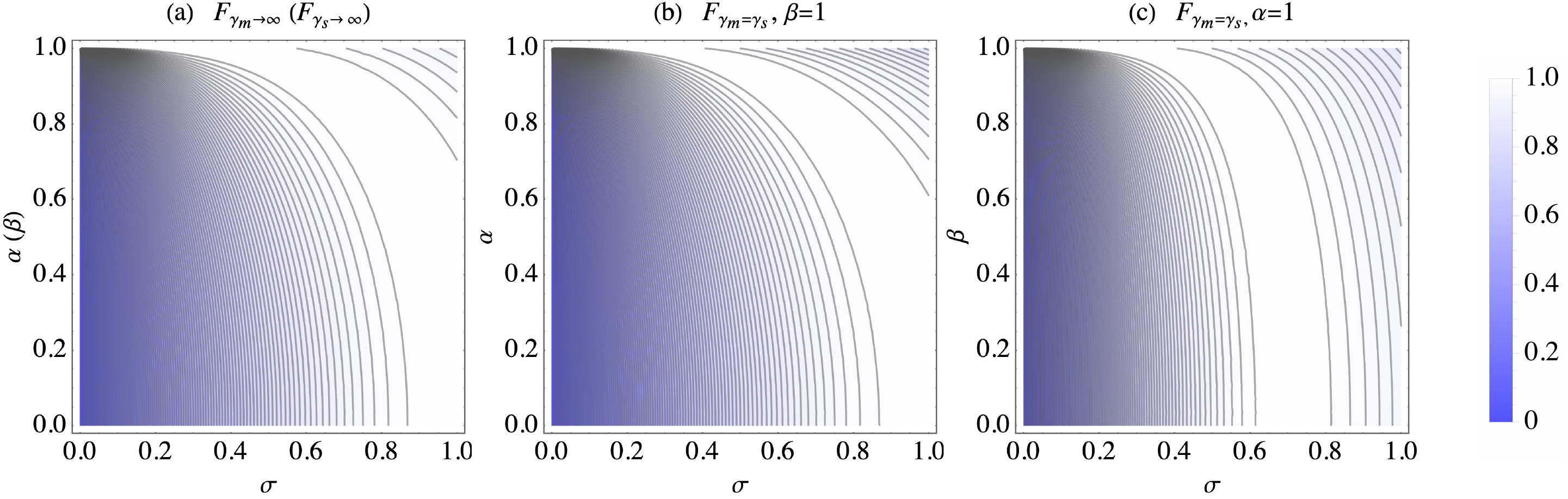}
  \caption{Behavior of the teleportation fidelity for the different cases described in the text. Plot (a) shows the behavior with $\alpha$, the state entanglement, and $\sigma= \gamma_c/\gamma_s$, for the ideal SFG measurement, with $\beta = 1$ and $\gamma_m \rightarrow \infty$, as considered in Ref \cite{Humble2010}. The same plot describes the fidelity as a function of $\beta$ and $\sigma = \gamma_c/\gamma_m$ for the case of a maximally entangled state with $\alpha =1$ and $\gamma_s\rightarrow \infty$. Plots (b) and (c) illustrate the behavior of the fidelity when the entangled state and the entangled measurement have comparable bandwidths (here $\gamma_s = \gamma_m=1$). Here the fidelity behaves differently with $\alpha$ and with $\beta$, because $f_s$ and $f_m$ are not in general interchangeable in the expression for $\psi_{b|m}$. All quantities are dimensionless.}
  \label{fig:fidt}
\end{figure*}

We point out here that we have centered both $f_s$ and $f_m$ at 0 in frequency space, without loss of generality. This is because, in the protocol described in reference \cite{Humble2010}, Alice communicates her obtained frequency $\omega_a+\omega_c$ to Bob, whereupon he performs the appropriate frequency translation to his photon $b$ to recover the state that would have resulted, had Alice obtained $\omega_a+\omega_b$ in her measurement. Further note that we are using the parameters $\alpha$ and $\beta$ to quantify the entanglement of the shared state and the joint measurement, respectively, rather than a more familiar measure of entanglement for pure states, such as the Schmidt number \cite{Parker2000}. We have made this choice because, although the Schmidt number $K$ bears a simple relationship with our parameter $\alpha$ (or $\beta$), satisfying $K=\frac{1}{\sqrt{1-\alpha^2}}$ (see Appendix \ref{app:schmidt}), the latter has the convenient feature of being bounded by the interval $[-1,1]$, whereas the Schmidt number diverges for maximal entanglement. 

With all of this in consideration, Alice's joint measurement on photons a and c heralds Bob's photon b in the teleported state

\begin{equation}
\label{psibm}
\begin{gathered}
    \ket{\psi_{b|m}} = \int \dw_b \psi_{b|m}(\omega_b)\adag_b(\omega_b)\vac,\\
    \psi_{b|m}(\omega_b) = N_{b|m}\int \dw_a \dw_c f^*_m(\omega_a, \omega_c)f_s(\omega_a,\omega_b)\psi_c(\omega_c).
\end{gathered}
\end{equation}
where $N_{b|m}$ is the appropriate normalization constant. The teleportation fidelity is then given by the modulus squared of the overlap,
\begin{equation}
    F=\left|\braket{\psi_c|\psi_{b|m}}\right|^2 = \left|\int \dw \psi_c^*(\omega)\psi_{b|m}(\omega)\right|^2
\end{equation}
For this analysis, we let $\psi_c$ be a Gaussian function with characteristic width $\gamma_c$,

\begin{equation}
    \psi_c(\omega)= \frac{1}{\sqrt{\gamma_c \sqrt{\pi}}} e^{-\omega^2/2\gamma_c^2}.
\end{equation}

Using this form for the states and measurements, we obtain an algebraic expression for the fidelity which depends on five parameters, $F = F(\alpha,\beta,\gamma_s,\gamma_m,\gamma_c)$. The full expression is unwieldy and not very instructive to display here. We shall verify that our formalism reproduces the result of reference \cite{Humble2010} in the appropriate limits. That reference studies the behavior of the fidelity as a function of $\alpha$ and $\sigma=\gamma_c/\gamma_s$ for a uniformly phasematched SFG process followed by an ideally-resolved frequency detection. This corresponds to taking the limit $\gamma_m \rightarrow \infty$ and $\beta =1$. In these limits, our formalism exactly recovers the fidelity

\begin{comment}
\begin{equation}
\begin{gathered}
F(\alpha, \beta=1,\gamma_s=1,\gamma_m\rightarrow\infty,\gamma_c=\sigma)\\
=\sqrt{\frac{4 \sigma^2 (\sigma^2+1)(\sigma^2+1-\alpha^2)}{((\sigma^2+1)^2-\alpha^2)^2}},
\end{gathered}
\end{equation}
\end{comment}

\begin{equation}
\begin{gathered}
F_{\gamma_m\rightarrow\infty}=\sqrt{\frac{4 \sigma^2 (\sigma^2+1)(\sigma^2+1-\alpha^2)}{((\sigma^2+1)^2-\alpha^2)^2}},
\end{gathered}
\end{equation}
which is displayed in Fig. \ref{fig:fidt} (a). In that reference, an interesting feature of this behavior of the fidelity was noted. That is, although the fidelity increases monotonically with the source entanglement $\alpha$ for $\sigma \ll 1$, this is no longer true for when $\gamma_c$ is comparable to $\gamma_s$. In particular, the fidelity is equal to one along the curve $\alpha^2 = 1-\sigma^4$, and is equal to $\sqrt{8/9}$ at the upper-right hand corner of the plot, where $\alpha=1$ and $\sigma=1$. In the language of our formalism, given the ideal entangled measurement, with infinite SFG bandwidth and ideal spectral resolution, there is a trade-off between spectral bandwidth and spectral entanglement of the sources.

Our result allows us to generalize further, however, and also consider the case of the Gaussian SFG measurement with finite bandwidth. First we consider the reverse scenario to the one above, where the source is perfectly entangled, with $\gamma_s \rightarrow \infty$ and $\alpha=1$, and look at the dependence of the fidelity on $\beta$ and $\sigma$. In this case we find that the fidelity exhibits the same dependence, that is, 
%$F(1,\beta,\gamma_s\rightarrow\infty,\gamma_m=1,\sigma)=F(\beta,1,\gamma_s=1,\gamma_m\rightarrow\infty,\sigma)$. Again, this fidelity is maximized for $\beta^2=1-\sigma^4$, 

\begin{equation}
F_{\gamma_s\rightarrow\infty}=\sqrt{\frac{4 \sigma^2 (\sigma^2+1)(\sigma^2+1-\beta^2)}{((\sigma^2+1)^2-\beta^2)^2}},
\end{equation}
and we can conclude that, given an ideal entangled state between Alice and Bob, there is a trade off between spectral bandwidth and spectral entanglement of the measurement.

Finally, we arrive at the most realistic case, where both the entangled source and the measurement have finite bandwidths, corresponding to finite phasematching in the PDC and SHG processes. Here we set them equal, taking $\gamma_s = \gamma_m=1$, and obtain

\begin{equation}
\begin{gathered}
F_{\gamma_m=\gamma_s}=\sqrt{\frac{4 \sigma^2 (\beta^2-2(1+\sigma^2))(\beta^2-(2-\alpha^2)(1+\sigma^2))}{(1+\sigma^2)^2(\alpha^2+\beta^2-2(1+\sigma^2))^2}}.
\end{gathered}
\end{equation}
In this case we find the interesting and counterintuitive result that the behaviors of the fidelity with the source entanglement $\alpha$ and with the measurement entanglement $\beta$ are no longer equivalent. We show this by plotting the behavior of the limiting cases of $F_\gamma(\alpha,1,\sigma)$ (spectral resolution of the SFG) and $F_\gamma(1,\beta,\sigma)$ (monochromatic pumping of the PDC) in Fig. \ref{fig:fidt} (b) and (c), respectively. In the case of $\beta=1$, the fidelity is maximized along the curve $\alpha^2  = \frac{1+\sigma^2 - 2\sigma^4}{1+\sigma^2}$ and has similar limiting behaviors to the ideal case considered in reference \cite{Humble2010}. The case of $\alpha = 1$ exhibits a starker contrast, taking its maximum value along the curve $\beta^2=\frac{-1+\sigma^2+2\sigma^4}{-1+\sigma^2}$. Unlike any of the previous cases, the fidelity is no longer equal to unity in the bottom right-hand corner, for $\sigma = 1$, $\beta = 0$, but instead it is equal to $\sqrt{8/9}$.

%The main discrepancy between the two plots is an additional reduction of the fidelity near $\sigma =1$ for the case of $\gamma_m=1$. This can be understood as follows: in both cases, the finite bandwidth of the PDC state JSA acts as a Gaussian filter on the teleported state, which explains the reduction of the fidelity when $\sigma$ is comparable to $\gamma_s$. However, when in addition, the SFG measurement JSA is finite, this filtering happens twice, and the reduction is compounded.

%We plot the fidelity as a function of $\alpha$ and $\beta$ in figure \ref{fig:fid}. It can be seen readily that the fidelity is maximized when not only the shared $a$ and $b$ two-photon state is highly entangled ($\alpha \rightarrow 1$), but in addition the \textit{measurement} on photons $a$ and $c$ is highly entangled ($\beta \rightarrow 1$). The unit fidelity in the limit $\alpha = 1$ is consistent with the case considered in reference \cite{Molotkov1998}, i.e., where $f_s(\omega_a,\omega_b) \rightarrow \delta(\omega_a+\omega_b)$, which corresponds to a PDC source pumped with a monochomatic, rather than pulsed, laser. However, the limit $\beta=1$ tells us that, additionally, to achieve maximal teleportation fidelity with arbitrary input spectra, the SFG photon must be spectrally resolved, leading to $f_m(\omega_a,\omega_c) \rightarrow \delta(\omega_a+\omega_c)$. 

We emphasize that $\beta < 1$ does not represent a non-ideal spectral resolution of the upconverted photon, since we are only considering projective measurements, but instead corresponds to a coherent broadband measurement, as could be obtained using a quantum pulse gate. What this last result suggests is that, for finite bandwidths of the entangled source and the entangled measurement, it is not generally the case that spectral resolution maximizes the teleportation fidelity. Further, the asymmetry between the behaviors of entangled state and the entangled measurement can be understood from the fact that the state JSA $f_s$ and the measurement JSA $f_m$ are not interchangeable in the expression for $\psi_{b|m}$, with $f_m$ having both of its arguments integrated over. Most notably, we have shown that, by treating two-photon measurements more generally and on equal footing with the two-photon states, it is possible not only to recover previously-obtained results in the limit of ideal measurements, but also to uncover which states and measurements are optimal for a given task (in this case spectral teleportation), under more realistic constraints (in this case, finite PDC and SFG bandwidths).

This brief analysis leaves open the question of how to generalize to a more realistic, non-ideally resolved SFG measurement. For a mixed bipartite state $\hat{\rho}$, a convenient measure of entanglement is the \textit{negativity} \cite{Vidal2002}. The negativity essentially counts the negative eigenvalues of $\hat{\rho}$ partially transposed with respect to one of its subsystems, and it sets an upper bound on the teleportation capacity of the state. This suggests that we may define a negativity associated with a non-projective POVM element $\hat{\Pi}_q$ as the negativity of its mixed retrodicted state $\hat{\rho}_q$. The role of finite spectral resolution in SFG detection has been investigated numerically for entanglement swapping in reference \cite{Vitullo18}. However, it could be more elegant to frame this relationship in terms of the negativities both of the input states and the measurements in scenarios such as quantum teleportation and entanglement swapping, and this remains to be explored in future work.\\

%%%%%%%%%%%%%%%%%%%%%%%%%%%%%%%%%%%%%%%%%%%%%%%%%%%%%%%%%%%%%%%%
\section{Conclusion}
We have demonstrated how to construct the POVM associated with two-photon detection by SFG followed by temporal-mode-selective single-photon detection. We have shown that this POVM is proportional to the two-photon state created in the time-reverse PDC process pumped with a field in the detected mode. This allowed us to characterize several aspects of the POVM relevant to its adequacy for quantum information protocols. In particular, we have shown that a projective measurement of the SFG photon corresponds to a projective two-photon POVM element. We have pointed out the special case where orthogonal SFG single-photon measurements correspond to orthogonal two-photon measurements. And finally, we have shown the correspondence between the two-photon entanglement retrodicted by the SFG measurement and the two-photon entanglement produced by the time-reversed PDC process. These results could have implications for quantum information experiments relying on PDC and SFG in terms of exploring the interplay between entangled states and entangled measurements. Additionally, it remains an open question how best to \textit{certify} the entanglement of the SFG measurement \cite{Bennet2014}, or even to perform quantum tomography of the process. Finally, given recent interest in using quantum light for two-photon absorption \cite{Schlawin2017} \cite{landes2020}, our results open the question of whether it's possible to have a combined framework of two-photon processes in terms of quantum measurement theory.

\begin{acknowledgments}
We would like to acknowledge M. G. Raymer and S. J. van Enk for valuable discussions. This work is funded by NSF grant No. 1839216.
\end{acknowledgments}

\section*{Appendix}

\begin{appendix}

\section{Deriving the three-wave mixing transformation} \label{app:ham}

Strictly speaking, the Hamiltonian describing the nonlinear interactions we consider is a time-dependent quantity, $\hat{H}(t)$, whereby a state $\ket{\Psi_\text{out}}$ evolves from an initial state $\ket{\Psi_\text{in}}$ according to

\begin{equation}
\begin{gathered}
	\ket{\Psi_\text{out}}= \exp{\Big[-\frac{i}{\hbar}\int_0^t \text{d}t' \hat{H}(t')\Big]}\ket{\Psi_\text{in}}\\
	\approx \left(1 - \frac{i}{\hbar}\int_0^t \text{d}t' \hat{H}(t')\right)\ket{\Psi_\text{in}}
	\label{eq:outstate}
\end{gathered}
\end{equation}
The relevant Hamiltonian for three-wave mixing has the form

\begin{equation}
	\hat{H}(t) = \chi \int_V \text{d}V \hat{E}^+_p(\mathbf{r},t)\hat{E}^-_s(\mathbf{r},t)\hat{E}^-_i(\mathbf{r},t) + \text{H.c.}
\end{equation}
where $\hat{E}^{+(-)}_j$ denotes the positive (negative) frequency component of the $j$ field operator, with $j=p,s,i$. $V$ denotes the interaction volume, which we take to be infinite in the transverse direction (by assuming the field modes are well-confined within the crystal area), and of length $L$ in the longitudinal direction. Finally, $\mathbf{r}$ and $t$ denote the space and time coordinates, and $\tilde{\chi}$ describes the interaction strength. We expand the field operators into their plane-wave components,

\begin{equation}
\begin{gathered}
	\hat{E}^+_j(\mathbf{r},t)=\int \text{d}\omega_j A_j(\omega_j)\exp{\Big[i(\mathbf{k}_j(\omega_j) \cdot \mathbf{r} - \omega_j t)\Big]} \hat{a}_j(\omega_j),\\
	\hat{E}^-_j = (\hat{E}^+_j)^\dagger,
\end{gathered}	
\end{equation}
where $A_j(\omega_j)$ is a slowly-varying function of $\omega$. Substituting these into the Hamiltonian and absorbing all the slowly-varying functions into $\chi$, we obtain

\begin{align}
	\hat{H}(t) = & \chi \int_V \text{d}V \int \dw_p\dw_s\dw_i \hat{a}_p(\omega_p)\adag_s(\omega_s)\adag_i(\omega_i)\\
	\nonumber & \times \exp\Big[i(\mathbf{k}_p(\omega_p)-\mathbf{k}_s(\omega_s)-\mathbf{k}_i(\omega_i))\cdot\mathbf{r}\Big] \\
	\nonumber & \times \exp\Big[-i(\omega_p-\omega_s-\omega_i)t\Big] + \text{H.c.}.
\end{align}

Now we use this form of the Hamiltonian to compute output state \eqref{eq:outstate} to first order in the expansion, whereupon we carry the integration over the transverse spatial directions to infinity. Additionally, we carry out the time integral from negative to positive infinity because the input and output states are observed long before and after the interaction time $t$, resulting in a delta-function in $(\omega_p-\omega_s-\omega_i)$ (energy conservation). All of this obtains

\begin{equation}
\begin{gathered}
	\ket{\Psi_\text{out}}  \approx \Bigg[1-i\chi \int_0^L \text{d}z \int \dw_s\dw_i \exp\Big[i(\Delta\mathbf{k})_z z\Big]\\
	\times \hat{a}_p(\omega_s+\omega_i)\adag_s(\omega_s)\adag_i(\omega_i)+\text{H.c.}\Bigg]\ket{\Psi_\text{in}},
\end{gathered}
\end{equation}
where we have also absorbed the $\hbar$ into $\chi$. Carrying out the integration over $z$ provides the phase-matching function $\Phi(\omega_s,\omega_i)$, and we define the transformation

\begin{equation}
	\hat{H}=\chi \int \dw_s\dw_i \Phi(\omega_s,\omega_i) \hat{a}_p(\omega_s+\omega_i)\adag_s(\omega_s)\adag_i(\omega_i) + \text{H.c},
\end{equation}
such that
\begin{equation}
	\ket{\Psi_\text{out}} \approx \left(1 - i\hat{H}\right)\ket{\Psi_\text{in}}.
\end{equation}

\section{Relating the entanglement parameter $\alpha$ to the Schmidt number $K$} \label{app:schmidt}

In section \ref{sec:entanglement} we used the scenario of spectral teleportation to illustrate the role of entanglement in the measurement, on par with entanglement in the state, in a quantum protocol. To that end, we quantified the teleportation fidelity in terms of the correlation parameters $\alpha$ ($\beta$) of the bivariate Gaussian state $f_s(\omega,\omega')$ (measurement $f_m(\omega,\omega')$). This parameter has the advantage of being bounded by the interval $[-1,1]$, with maximal entanglement at the boundaries, whereas more common measures of entanglement for pure states, such as the entropy and the Schmidt number, diverge for maximal entanglement. Here we show for completeness how the Schmidt number $K$ depends functionally on $\alpha$, while the same analysis holds for $\beta$. 

The Gaussian JSA $f_s(\omega,\omega')$ from \eqref{eq:fstate} has a Schmidt decomposition of the form

\begin{equation}
	f_s(\omega,\omega') = \sum^\infty_{j=0}\ \sqrt{\lambda_j}\ u_j(\omega) v_j(\omega'),
\end{equation}
where $\{u_j(\omega)\}$ is the orthonormal set of Hermite-Gauss functions spanning the spectral Hilbert space over $\omega$, and the same is true of $\{v_j(\omega')\}$ \cite{Humble2010}. The Schmidt coefficients $\lambda_j$ are given by

\begin{equation}
	\lambda_j=\text{sech}^2\ \zeta\ \text{tanh}^{2j}\ \zeta,
\end{equation}
satisfying $\sum_{j=0}^\infty\lambda_j = 1$, and where $\zeta$ is given by

\begin{equation}
	\alpha = \tanh\ 2\zeta.
	\label{eq:alpha}
\end{equation}

The Schmidt number $K$ is then given by

\begin{equation}
	K = \frac{1}{\sum_{j=0}^\infty \lambda_j^2} = \text{cosh}\ 2\zeta.
	\label{eq:schmidt}
\end{equation}
Combining Eq. \eqref{eq:alpha} and \eqref{eq:schmidt}, we arrive at the simple relationship

\begin{equation}
	K = \frac{1}{\sqrt{1-\alpha^2}},
\end{equation}
where, as expected, $K$ is equal to unity for the case of no correlation, $\alpha = 0$, and diverges for maximal correlation, $\alpha=\pm1$.

\end{appendix}

\bibliography{biblio}

\end{document}